\documentclass[runningheads,a4paper]{llncs}
\usepackage{graphicx,natbib,amssymb,lineno,amsmath,latexsym,wrapfig}

\makeatletter
\def\elsartstyle{%
    \def\normalsize{\@setfontsize\normalsize\@xiipt{14.5}}
    \def\small{\@setfontsize\small\@xipt{13.6}}
    \let\footnotesize=\small
    \def\large{\@setfontsize\large\@xivpt{18}}
    \def\Large{\@setfontsize\Large\@xviipt{22}}
    \skip\@mpfootins = 18\p@ \@plus 2\p@
    \normalsize
}
\@ifundefined{square}{}{}
\makeatother

\newcommand{\ket}[1]{\ensuremath{|#1\rangle}}

\newcommand\remove[1]{}
\newcommand{\word}[1]{\emph{#1}}

\begin{document}
\title{Extracting Spooky-activation-at-a-distance from Considerations of Entanglement}

\author{Peter Bruza\inst{1} \and Kirsty Kitto\inst{1} \and
Douglas Nelson\inst{2}\and Cathy McEvoy\inst{3}}

\institute{Faculty of Science and Technology, Queensland University of Technology\\\email{p.bruza@qut.edu.au, kirsty.kitto@qut.edu.au}\and Department of Psychology, University of South Florida\\\email{dneslon@cas.usf.au} \and School of Ageing Studies, University of South Florida \\\email{cmcevoy@cas.usf.edu}}

\maketitle

\begin{abstract}
Following an early claim by Nelson \& McEvoy \cite{Nelson:McEvoy:2007} suggesting that word associations can display `spooky action at a distance behaviour', a serious investigation of the potentially quantum nature of such associations is currently underway. This paper presents a simple quantum model of a word association system. It is shown that a quantum model of word entanglement can recover aspects of both the Spreading Activation model and the Spooky model of word association experiments.
\end{abstract}

\section{Modelling Words and Meaning}

Human beings are adept and drawing context-sensitive associations and inferences across a broad range
of situations ranging from the mundane to the creative inferences that lead to scientific discovery. Such
reasoning has a strong pragmatic character and is transacted with comparatively scarce cognitive assets. However, despite our apparent proficiency at drawing inferences, and our ability to express words in such a manner that other people can (usually) understand the meaning that we are trying to convey, our theoretical understanding of how this process occurs has been slow to develop. 

   The field of cognitive science has recently produced an ensemble of semantic models which have an
encouraging, and at times impressive track record of replicating human information processing, such as
human word associations norms \cite{Article:96:Burgess:Cognition,Article:98:Burgess:HAL,Article:00:Lowe:SemanticSpace,Article:01:Lowe:SemanticSpace,Article:97:Landauer:LSA,Article:98:Landauer:LSA,Article:97:Patel:SemanticSpace,Article:99:Levy:SemanticSpace,Article:02:Sahlgren:SemanticSpace}. The term ``semantic'' derives from the
intuition that words seen in the context of a given word contribute to its meaning, or, more colloquially expressed,
the meaning of a word is derived from the ``company it keeps'' \cite{Article:01:Kintsch:LSA}. 
In order to progress in our understanding of how meaning is generated from sets of words in a language we must understand the way in which the mental lexicon of that language is generated during language acquisition, and how it works once created in the mind of a specific individual.

\subsection{The Mental Lexicon}

The mental lexicon of a language refers to the words of a language, but its structure is represented by the associative links that bind this vocabulary together.  Such links are acquired through
experience and the vast and semi-random nature of this experience
ensures that words within this vocabulary are highly
interconnected, both directly and indirectly through other words.
For example, during childhood development and the associated acquisition of English, the word \word{planet} becomes associated with
\word{earth}, \word{space}, \word{moon}, and so on. Even within
this set, \word{moon} can itself become linked to \word{earth} and
\word{star} etc.  Words are so associatively interconnected with each other
that they meet the qualifications of a `small world' network
wherein it takes only a few associative steps to move from any one
word to any other in the lexicon \cite{Steyvers:Tenenbaum:2005}.
Because of such connectivity
individual words are not represented in long-term memory as
isolated entities but as part of a network of related words.
However, depending upon the context in which they are used, words
can take on a variety of different meanings and this is very
difficult to model \cite{gabora:rosch:ea:toward}.

Much evidence shows that for any individual, seeing or hearing a word activates words related to it through prior learning.  Understanding how such activation affects memory requires a map of links among known words, and free association provides one reliable means for constructing such a map \cite{Nelson:McEvoy:Schreiber:2004}. In free association experiments, words are presented to large samples of participants who produce the first associated word to come to mind.  The probability or strength of a pre-existing link between words is computed by dividing the production frequency of a response word by its sample size.  For example, the probabilities that \word{planet} produces \word{earth} and \word{mars} are 0.61 and 0.10, respectively, and we say that \word{earth} is a more likely or a stronger associate of \word{planet} than \word{mars}.  

Just like the nonlocality experiments of quantum theory, human memory experiments require very careful preparation of the state to be tested.  For example, in extralist cuing, participants typically study a list of to-be-recalled target words shown on a monitor for 3 seconds each (e.g., \word{planet}).  The study instructions ask them to read each word aloud when shown and to remember as many as possible, but participants are not told how they will be tested until the last word is shown.  The test instructions indicate that new words, the test cues, will be shown and that each test cue is related to one of the target words just studied (e.g., \word{universe}).  These cues are not present during study (hence, the name extralist cuing).  As each cue is shown, participants attempt to recall its associatively related word from the study list. In contrast, during intralist cuing the word serving as the test cue is presented with its target during study (e.g., \word{universe planet}). Participants are asked to learn the pairing, but otherwise the two tasks are the same. It appears that more associate-to-associate links benefit recall, and two  competing explanations for this phenomenon have been proposed: Spreading Activation, and Spooky Activation At a Distance.

This paper will demonstrate an intriguing connection between these two explanations, obtained by making the assumption that words can become entangled in the human mental lexicon. 

\subsection{Isn't Entanglement Correlation?}

Entanglement is a phenomenon unique to quantum behaviour. If a system consisting of two components becomes entangled then it cannot be thought of as separate anymore; a description of one component without reference to the other will, in some cases, fail. Indeed, an entangled quantum system will generally exhibit an intercomponent agreement with reference to any combination of measurement settings. This is of particular importance for a system that becomes spatially extended, as in the case where the two components are taken a long way from each other. Here we find that quantum systems display correlation instantaneously in response to what might even be a delayed choice of measurement setting \cite{aspect.dalibard.ea:experimental}, and yet cannot be used to transmit information between two observers, and thus does not actually violate Special Relativity \cite{maudlin:quantum}. 
This is in contrast with classical scenarios of correlation. In a classical situation a system is in a pre-existing state, and this is discovered through the process of measurement. Not so with a quantum system, where the process of measurement can actively influence the outcome itself. This fundamental difference between the two types of system was first alluded to in the by now famous EPR\footnote{Einstein--Podolsky--Rosen} debate, but was only inescapably highlighted with the more subtle (and recent) results surrounding the contextuality of quantum systems (see  \cite{laloe:do} or \cite{ballentine:quantum} for a good introduction to these ideas). An entangled quantum system is very different from a correlated classical system; no pre-existing elements of reality \cite{einstein.podolsky.ea:can} have been found that can explain the agreement that is obtained between distant measuring devices that are set to determine the state of a quantum system.

To make these ideas more concrete, let us consider a specific example of classical correlation.
If the same number
is written on two pieces of paper, enclosed in two envelopes, and sent
to Alice and Bob at two distant ends of the Universe, the information obtained upon opening of
one of the envelopes will instantly correlate with the state of the other
envelope at the other end of the universe. However, these correlated pieces of paper are not entangled. The number on the two pieces of paper can be regarded as a hidden variable, or element of reality; even before we open the envelope it exists, in both envelopes. Upon opening the envelope at one end of the Universe we find out what that number is, and hence know what number is already inscribed upon the other piece of paper. The quantum analogue of this scenario would be far stranger. The situation most similar to the nonlocal effects exhibited by entangled quantum systems would involve Alice, at one end of the Universe choosing to write a number upon her blank piece of paper when she opens her envelope, and then finding that Bob, upon opening his envelope found exactly the same number upon his piece of paper at the other end of the Universe. Obviously this does not happen. 

However, we might ask if similar cases of intercomponent dependency, or spooky-activation-at-a-distance, exist for systems beyond the field of physics.

We shall now look at the problem of modelling associate-to-associate links in the human mental lexicon, before showing how the assumption that associates might be entangled in a subject's cognitive state can lead to a new model of word associations.

\section{Modelling Associate-to-associate Links}

\begin{figure}
\begin{center}
\includegraphics[height=5cm]{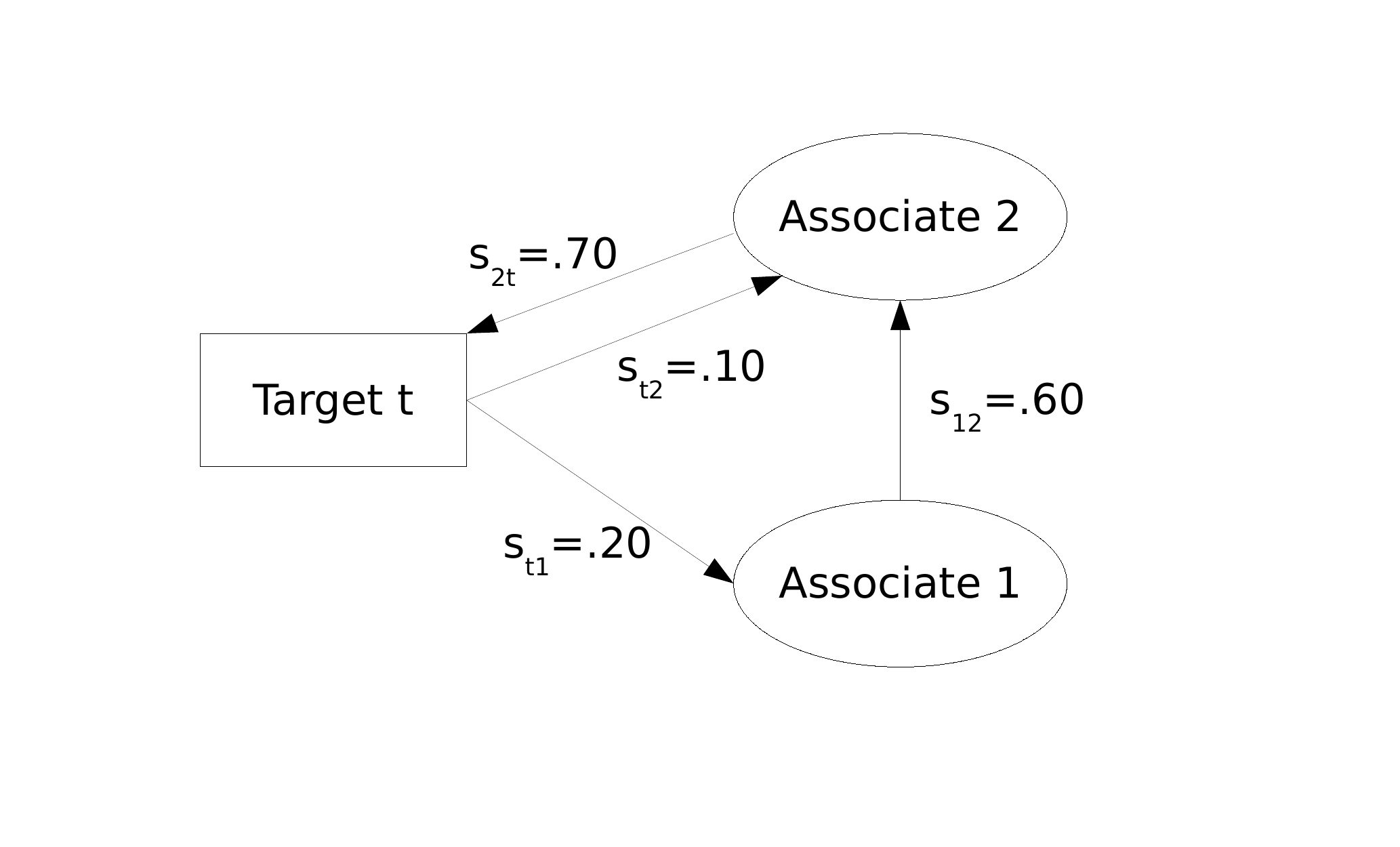}
\caption{A hypothetical target with two associates and single associate-to-target and associate-to-associate links. From Nelson, McEvoy, and Pointer
(\cite{Nelson:McEvoy:Pointer:2003}).}
\label{target-t}
\end{center}
\end{figure}

Figure \ref{target-t} shows a hypothetical target word having two target-to-associate links in a subject's cognitive state.  
There is also an associate-to-associate link between Associates 1 and 2, and an associate-to-target link from Associate 2 to the Target $t$.  The values on the links indicate relative strengths estimated via free association. Nelson et al.  have investigated reasons for the more likely recall of words having more associate-to-associate links \cite{Nelson:McEvoy:Pointer:2003}.  Two competing explanations for why associate-to-associate links benefit recall have been proposed.

The first is the  Spreading Activation equation, which is based on the classic idea that activation spreads through a fixed associative network, weakening with conceptual distance (e.g., \cite{Collins:Loftus:1975}):
\begin{align} \label{eq:spreadact}
S(t) & =  \sum_{i=1}^nS_{ti}S_{it} + \sum_{i=1}^n\sum_{j=1}^nS_{ti}S_{ij}S_{jt} \\
       & =  (.10 \times .70) + (0.20 \times 0.60 \times 0.70) \\
       & =  0.154
\end{align}
where $n$ is the number of associates and $i \neq j$.
$S(t)$ denotes the strength of implicit activation of target $t$ due to study, $S_{ti}$ target-to-associate activation strength, $S_{it}$ associate-to-target activation strength (resonance), and $S_{ij}$ associate-to-associate activation strength (connectivity).
Multiplying link strengths produces the weakening effect. Activation ostensibly travels from the target to and among its associates and back to the target in a continuous chain, and the target is strengthened by activation that returns to it from pre-existing connections involving two and three-step loops. More associate-to-associate links create more three-step loops and theoretically benefit target recall by increasing its activation strength in long-term memory. Importantly, note that the effects of associate-to-associate links are contingent on the number and strength of associate-to-target links because they allow activation to return to the target. If associate-to-target links were absent, even the maximum number of associate-to-associate links would have no effect on recall because activation could not return to the target.

In contrast, in the `Spooky Activation at a Distance' equation, the target activates its associative structure in synchrony:
\begin{align}\label{eq:spooky}
S(t) & =  \sum_{i=1}^nS_{ti} + \sum_{i=1}^nS_{it} + \sum_{i=1}^{n}\sum_{j=1}^nS_{ij} \\
        & =   0.20 + 0.10 + 0.70 + 0.60  \\
        & =   1.60
\end{align}
where $i \neq j$;  $S_{ti}$, target-to-associate $i$ strength;  $S_{it}$, associate $i$-to-target strength (resonance);  $S_{ij}$, associate $i$-to-associate $j$ strength (connectivity)

This equation assumes that each link in the associative set contributes additively to the target's net strength.  The beneficial effects of associate-to-associate links are not contingent on associate-to-target links. Stronger target activation is predicted when there are many associate-to-associate links even when associate-to-target links are absent. In fact, associate-to-target links are not special in any way. Target activation strength is solely determined by the sum of the link strengths within the target's associative set, regardless of origin or direction.

\section{Entanglement of Words}

How should we represent the combination of words in the human mental lexicon? QT uses the tensor product, $\otimes$, to denote composite systems. Consider the case of $m=2 $ study words: $u$ and $v$ presented to a group of subjects. Let us assume that, when cued, the subjects recall neither target word. In this case we could write:
\begin{equation}
 \ket{u}\otimes\ket{v}=\ket{0}\otimes\ket{0} = \ket{00}
\end{equation}
where the notation $\ket{00}$ is just shorthand for the tensor product state $\ket{0}\otimes\ket{0}$ describing the composite system of two negative outcomes.
If word $u$ alone was recalled then we would write $ \ket{u}\otimes\ket{v}=\ket{10}$, whereas in the converse case we would write $ \ket{01}$ and finally, if both words were recalled then the tensor product would yield the state $\ket{11}$.

However, this straightforward scenario is not the only form of situation possible in the quantum formalism. Superposition states can also occur, and these are important as they can represent the situation where the words $u$ and $v$ may be more likely to be recalled in one context than another. Assume that we can represent one subject's cognitive state with reference to the combined targets $u$ and $v$ as a 2 $q$-bit register that refers to their states of `recalled' and `not recalled' in combination. Thus, if we represent the target words using the standard superpositions $\ket{u} = a_0\ket{0} + a_1\ket{1}$ and $\ket{v} = b_0\ket{0} + b_1\ket{1}$, (where $a_0^2+a_1^2=1$ and $b_0^2+b_1^2=1$), then it is possible to denote the state of the combined system by writing the tensor product
\begin{align}
 \ket{u}\otimes\ket{v} &=  \left(a_0\ket{0} + a_1\ket{1} \right) \otimes\left(b_0\ket{0} + b_1\ket{1} \right) \\
&= a_0b_0\ket{00} + a_1b_0\ket{10} + a_0b_1\ket{01} + a_1b_1\ket{11},\label{eq:producttwo}
\end{align}
where $ |a_0b_0|^2+|a_1b_0|^2+|a_0b_1|^2+|a_1b_1|^2=1$.
This is the most general state possible. It represents a quantum combination of the above four possibilities, obtained using a tensor multiplication between the states $\ket{u}$ and $\ket{v}$. In contrast to the simple cases discussed above, here no state of recall is `the' state, rather, we must cue the subject and elicit a response from them before we can talk about a word being `recalled' or `not recalled'. Indeed, a different cue might elicit a very different response, and the quantum formalism could deal with this via a change of basis. 

It is important to realise however, that \eqref{eq:producttwo} is \emph{not} the only form of state that can be obtained from combination of $\ket{u}$ and $\ket{v}$ in the quantum formalism.

The other form of state, an \emph{entangled state} is one that it is impossible to write as a product. 
As an example of an entangled state, we might consider the the state $\psi$ where the words  $u$ and $v$ are either \emph{both} recalled, or both \emph{not} recalled in relation to a cue $q$. One representation of this scenario is given by the following state:
\begin{equation}
\label{eqn:2bell+}
\psi  = \frac{1}{\sqrt{2}}(\ket{00} +  \ket{11}).
\end{equation}
This seemingly innocuous state is one of the so-called \emph{Bell states} in QT. It is impossible to write as a product state, thus it differs markedly from \eqref{eq:producttwo}.
The fact that entangled systems cannot be expressed as a product of the component states makes them non-separable.
More specifically, there are no coefficients which can decompose equation \eqref{eqn:2bell+} into a product state exemplified by equation \eqref{eq:producttwo} which represents the two components of the system, $u$ and $v$, as independent of one another. For this reason $\psi$ is not written as $\ket{u}\otimes \ket{v}$ as it can't be represented in terms of the component states $\ket{u}$ and $\ket{v}$.

\section{An Analysis of Spooky-activation-at-a-distance in Terms of Entanglement}
\begin{table}
\begin{center}
\begin{tabular}{|c|c|c|c|} \hline
           &$t$ & $a_1$ & $a_2$ \\ \hline
$t$      &      &  0.2     &  0.1  \\ \hline
$a_1$ &      &            & 0.6 \\ \hline
$a_2$ & 0.7 &           &       \\ \hline\hline
            &$p_t=0.7$ & $p_{a_1} = 0.2$ & $p_{a_2} = 0.7$ \\ \hline
\end{tabular}
\caption{Matrix corresponding to hypothetical target shown in Figure \ref{target-t}}
\label{tab:target-t}
\end{center}
\end{table}

Nelson and McEvoy have recently begun to consider the Spooky-activation-at-a-distance formula in terms of quantum entanglement, claiming that ``The activation-at-a-distance rule assumes that the target is, in quantum terms, entangled with its associates because of learning and practicing language in the world. Associative entanglement causes the studied target word to simultaneously activate its associate structure" \cite[p3]{Nelson:McEvoy:2007}.
The goal of this section is to formalise this intuition. At the outset, it is important that the quantum formalism be able to cater for the set size and connectivity effects described elsewhere \cite{bruza.kitto.ea:entangling}.
Recall  both set size and associative connectivity have demonstrated time and again robust effects on the probability of recall.
Because the  Spooky-activation-at-a-distance formula sums link strengths irrespective of direction, it encapsulates the idea that a target with a large number of highly interconnected associates will generate a high activation level during study.

Table \ref{tab:target-t} is a matrix representation of the associative network of the hypothetical target $t$ shown in Figure  \ref{target-t}.
The last line of the matrix represents the summation of free association probabilities for a given word. 
For example,
\begin{eqnarray} 
p_{a_2} &= &\Pr(a_2|t) + \Pr(a_2|a_1) \\ 
              &= & 0.1+ 0.6 \\
              &=& 0.7
\end{eqnarray}
These free association probabilities may be added as it is assumed that each free association experiment is independent, that is associate $a_2$ being recalled in relation to cue $t$ is assumed independent of it being recalled in relation to the cue $a_1$. Since free association experiments require a subject who has not been primed in any way beyond that provided by the cue, this appears to be a reasonable simplifying assumption. 

\begin{figure}
\begin{center}
\includegraphics[width=8cm]{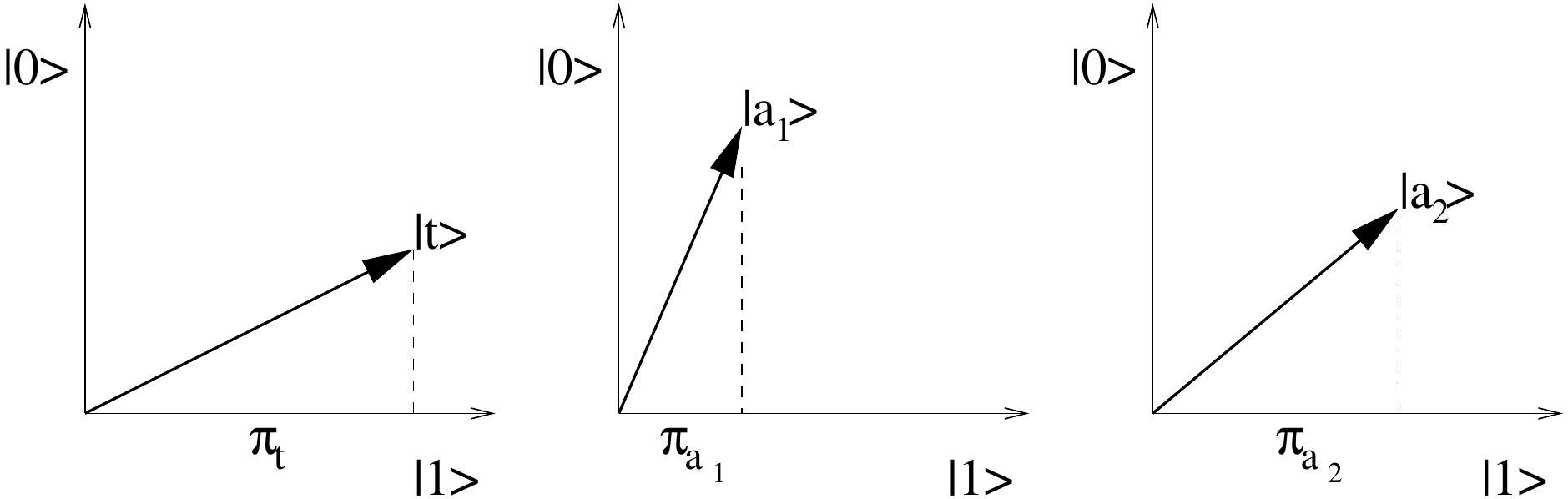}
\caption{Three bodied quantum system of words.}
\label{fig:qubit-register-t}
\end{center}
\end{figure}

Both Spreading Activation and Spooky-activation-at-a-distance approaches assume that free association probabilities determine the strength of activation of a word during study; they only differ in the way this activation strength is computed.
Viewing free association probabilities in this way allows the matrix to be considered as a many bodied quantum system modelled by three qubits. Figure~\ref{fig:qubit-register-t} depicts the system, here, each word is in a superposed state of being activated $\ket{1}$, or not $\ket{0}$. Note how each summed column with a non-zero probability leads to a qubit.

For ease of exposition in the following analysis, we shall change variables. The probabilities depicted in table~\ref{tab:target-t} are related to the probability densities of figure~\ref{fig:qubit-register-t} by taking their square root: e.g. $ \pi_t^2 = p_t$.
Using such a change of variables, the state of the target word $t$ would be written as:
\begin{equation}
\ket{t}  =  \sqrt{\bar{p_t}}\ket{0} + \sqrt{p_t}\ket{1}=\bar{ \pi_t}\ket{0}+ \pi_t\ket{1},
\end{equation}
where the probability of recall due to free association is $p_t= \pi_t^2$, and $\bar{p_t} = 1 - p_t=\bar{ \pi_t}^2$ represents the probability of a word not being recalled.
Thus, the states of the individual words are represented as follows in order to avoid cluttering the analysis with square root signs:
\begin{align}
\ket{t} & =  \bar{ \pi_t}\ket{0} +  \pi_t\ket{1} \label{eq:t} \\
\ket{a_1} &= \bar{ \pi}_{a_1}\ket{0} +  \pi_{a_1}\ket{1} \\
\ket{a_2} &= \bar{ \pi}_{a_2}\ket{0} +  \pi_{a_2}\ket{1} \label{eq:a2}
\end{align}
where $\bar{ \pi_t} = 1 -  \pi_t, \bar{ \pi}_{a_1} = 1 -  \pi_{a_1}$ and $\bar{ \pi}_{a_2} = 1 -  \pi_{a_2}$.

As detailed in the previous section, tensor products are used to model many bodied quantum systems.
The state $\psi_t$ of the most general combined quantum system is given by the tensor product of the individual states:
\begin{align}
\psi_t = & \ket{t} \otimes \ket{a_1} \otimes \ket{a_2} \\ 
         =  & (\bar{ \pi_t}\ket{0} +  \pi_t\ket{1}) \otimes (\bar{ \pi}_{a_1}\ket{0} +  \pi_{a_1}\ket{1}) \otimes (\bar{ \pi}_{a_2}\ket{0} +  \pi_{a_2}\ket{1}) \\
\begin{split}
= &\bar{ \pi_t}\bar{ \pi}_{a_1}\bar{ \pi}_{a_2}\ket{000} +
                \pi_t\bar{ \pi}_{a_1}\bar{ \pi}_{a_2}\ket{100} +
               \bar{ \pi_t} \pi_{a_1}\bar{ \pi}_{a_2}\ket{010} +
                \pi_t \pi_{a_1}\bar{ \pi}_{a_2}\ket{110} \\
             &  + \bar{ \pi_t}\bar{ \pi}_{a_1} \pi_{a_2}\ket{001} +
                \pi_t\bar{ \pi}_{a_1} \pi_{a_2}\ket{101} +
               \bar{ \pi_t} \pi_{a_1} \pi_{a_2}\ket{011} +
                \pi_t \pi_{a_1} \pi_{a_2}\ket{111}
\end{split}  \label{eq:productstate-t}
\end{align}

The intuition behind this expression is an enumeration of the possibilities of the states of the qubits. So, $\ket{111}$ represents the state in which all respective qubits collapse onto their state $\ket{1}$. 
In other words, $\ket{111}$ denotes the state of the system in which words $t$, $a_1$ and $a_2$ have all been \emph{activated} due to study of target $t$.
The probability of observing this is given by the taking the square of the product $ \pi_t \pi_{a_1} \pi_{a_2}$.
Conversely, the state $\ket{000}$ corresponds to the situation in which none of the words have been activated.

The state $\psi_t$ of the three-bodied system does not capture Nelson \& McEvoy's intuition that the studied target word $t$ simultaneously activates its associative structure. Recall their suggestion that target $t$ activates its associates in synchrony:  When target $t$ is studied it activates \emph{all} of its associates, or none at all.
Interestingly, this idea is quite easily captured by the assumption that the state $\psi_t$ evolves into a Bell entangled state $\psi'_t$ of the form:
\begin{equation}
\psi'_t = \sqrt{p_0}\ket{000} +\sqrt{p_1}\ket{111}.
\end{equation}
This formula represents a superposed state in which the entire associative structure of a subject is in a state of potential activation ($\ket{111}$) or potential non-activation ($\ket{000})$, with the probabilities of these states occurring given by $p_0$ and $p_1$ respectively. 

How should we ascribe values to the probabilities $p_0$ and $p_1$?
In QT these values would be determined by the unitary dynamics evolving $\psi_t$ into $\psi'_t$.
As such dynamics are yet to be worked out for cognitive states, we are forced to speculate. 
One approach is to assume the lack of activation of the target is determined solely in terms of lack of recall of any of the associates. That is,
\begin{equation}
p_0= \bar{p_t}\bar{p}_{a_1}\bar{p}_{a_2}
\end{equation}
Consequently, the remaining probability mass contributes to the activation of the associative structure as a whole. 
Thus, starting from the assumption of a Bell entangled state, we find that the probability $p_1$   expresses Nelson \& McEvoy's intuition:
\begin{align}
p_1 & =  1 - \bar{p_t}\bar{p}_{a_1}\bar{p}_{a_2} \\
       & = 1 - (1-p_t)(1-p_{a_1})(1-p_{a_2}) \label{eq:brackets} \\
       & = 1 - (1-p_t - p_{a_1} + p_tp_{a_1} - p_{a_2} + p_tp_{a_2}+ p_{a_1}p_{a_2} - p_tp_{a_1}p_{a_2}) \\
       & = \underbrace{p_t + p_{a_1} + p_{a_2}}_A + \underbrace{p_tp_{a_1}p_{a_2}}_B - \underbrace{(p_tp_{a_1}+p_tp_{a_2}+p_{a_1}p_{a_2})}_C \label{eq:terms}
\end{align}
Term $A$ corresponds to the summation of the free association probabilities in the above matrix. In other words, term $A$ corresponds exactly to the Spooky-activation-at-a-distance formula (See equation~\ref{eq:spooky}). 
At such, the assumption of a Bell entangled state provides partial support for the summation of free association probabilities which is embodied by the Spooky-activation-at-a-distance equation.
Ironically perhaps,  term $B$ corresponds to free association probabilities  multiplied according to the directional links in the associative structure. This is expressed in the second term of the spreading activation formula (See equation~\ref{eq:spreadact}).
In other words, departing from an assumption of entanglement leads to an expression of activation strength which combines aspects of both Spooky-activation-at-a-distance and Spreading Activation.

The third term $C$ is more challenging to interpret.
In a more complete model capable of generating some form of evolutionary dynamics we might expect that it would arise from the underlying structure of the Hilbert space used. Here, it has arisen from apparently sensible assumptions about how probabilities should be amassed in the the Bell entangled state. What significance can be drawn from this term? The beginnings of an answer can be formulated by returning to figure~\ref{target-t}. 
When seen in the context of actual values, the term $C$ has a significant compensating effect, subtracting 0.77 from the summation of the Spooky term with the spreading activation term:
\begin{align}
p_1 & =  A+B-C \\
\begin{split}
        & = (0.7+0.2+0.7)+(0.7\times0.2\times0.7) \\& \qquad \qquad - (0.7\times0.2+0.7\times0.7+0.2\times0.7) \\
\end{split}\\
       & = 0.928
\end{align}
Thus, according to this analysis, the strength of activation $p_1$ lies somewhere between Spreading Activation and Spooky-activation-at-a-distance.
Based on a substantial body of empirical evidence, Nelson and McEvoy have argued persuasively spreading activation \emph{underestimates} strength of activation \cite{nelson.mcevoy:entangled}. 
Here we have seen that when departing from an assumption that the associative structure is Bell entangled, a cautious preliminary conclusion is that Spooky-activation-at-a-distance overestimates the strength of activation and it is term $C$ which compensates for this.

We conclude this section with some remarks how the entanglement model above assists in the explanation of some experimental results that have not been well accounted for in current models of the mental lexicon. Attempts to map the associative lexicon soon made it clear that some words produce more associates than others.   This feature is called `set size' and it indexes a word's associative dimensionality \cite{Nelson:McEvoy:1979,Nelson:Schreiber:McEvoy:1992}. It was also revealed that the associates of some words are more interconnected than others.  Some words have many such connections (e.g., \word{moon-space}, \word{earth-planet}), whereas some have none, and this feature is called ``connectivity" \cite{Nelson:McEvoy:Pointer:2003}.  Thus, experiments have shown that in addition to link strengths between words,  the set size and connectivity of individual words have powerful effects on word recall, which existing theories cannot generally explain.

This is an interesting result for the model suggested here, as according to the above analysis we might surmise that the more associates a target has, the more qubits are needed to model it. When these are tensored the resulting space will have a higher dimensionality. Therefore a large set size is catered for by a tensor space of higher dimensionality.
Conversely, interconnectivity is catered for by larger probabilities in the initial superposed states of the respective qubits. Consider once again the matrix in table~\ref{tab:target-t}.
In the general case, when the associative structure is highly interconnected the sums of probabilities in the last row will tend to be higher. These will contribute to higher activation strength as they are summed  (as defined by probability $p_1$) in the same way as in Spooky-activation-at-a-distance. 
So it is possible to have a high activation strength  even though the tensor space has high dimensionality. 
When the associative structure in not interconnected these probabilities will be low and hence lead to lower strength of activation.

\section{Conclusions and Future Directions}

Obviously the model presented here is overly simple. The human mental lexicon is a vast and highly interconnected network consisting of thousands of words, and its associative structure is far more complex than the simple toy model in figure~\ref{target-t} could ever hope to represent. However, some promising initial results have been obtained from this very simplified analysis. We have seen some evidence that a quantum approach can model set size and connectivity effects, and a prediction has been made that Spooky-activation-at-a-distance overestimates associative behaviour by some factor. While equation~\eqref{eq:terms} makes a very concrete suggestion about how large this overestimation might be, it is unlikely that the situation will prove this simple.

Firstly, this is a very basic toy model, addressing the behaviour of only one target and two associates. A fully developed theory would have to be stated in terms of all the targets and all of their associates. It is very hard to extrapolate from this model to a more realistic one. One possibility would involve applying the apparatus of Statistical Mechanics in a density matrix approach to model sets of targets and their associates, but no results have been obtained here to date. 
A related issue surrounds the meaning of the term $C$ in equation~\eqref{eq:terms} as more associative structure is added. Here, we would see $C$ gaining more and more terms (since $A$ and $B$ essentially correspond to the two extreme values in the expansion of the bracketed term in \eqref{eq:brackets}). This does not seem plausible, and the way in which this term would act as it became larger is yet to be established.

In addition to these immediate issues, much work remains to be done in this area. Given the large amount of data collected about word association norms \cite{Nelson:McEvoy:Schreiber:2004} we might expect that experiments can be performed that might distinguish between the varying predictions of the different models, and work is underway to generate results here. A physically and cognitively motivated time evolution equation capable of generating the Bell-type state \eqref{eqn:2bell+} is essential before we can consider this model to be truly quantum(like) and initial ruminations about how this might be achieved are presented in \cite{kitto.bruza.ea:generalising}. 

Ending on a slightly more positive note, we consider the most important result of this article to be the indication that a straight tensorial combination of associate words in the mental lexicon is not particularly representative of the intuition that words and their associates are activated in synchrony. Given that a Bell-type entangled state provides a far more likely candidate for the behaviour of word association networks this result thus provides some first steps towards establishing evidence that human cognitive structures have some quantum(like) behaviour.

\subsubsection{Acknowledgements}

This project was supported in part by the Australian Research Council Discovery grant DP0773341 to P. Bruza and K. Kitto, and by grants from the National Institute of Mental Health to D. Nelson and the National Institute on Ageing to C. McEvoy.

\bibliographystyle{plain}

\end{document}